\documentclass[10pt]{iopart}
\usepackage{hyperref}
\usepackage{graphicx,color}
\usepackage[small,centerlast]{caption}
\usepackage{import}
\usepackage{siunitx}
\usepackage{bbm}

\usepackage[resetlabels]{multibib}
\newcites{SI}{nada}
\usepackage{amssymb}

\definecolor{mygrey}{rgb}{0.5,0.5,0.5}
\definecolor{mygreen}{rgb}{0,0.5,0}
\definecolor{mylightgreen}{rgb}{0,0.75,0}
\definecolor{myblue}{rgb}{0,0,0.75}
\definecolor{mymagenta}{cmyk}{.2,1,0,0.12}
\definecolor{myred}{rgb}{0.95,0,0}
\definecolor{myorange}{rgb}{1,.5,0}

\newcommand{\be}{\begin{equation}}
\newcommand{\ee}{\end{equation}}
\newcommand{\bea}{\begin{eqnarray}}
\newcommand{\eea}{\end{eqnarray}}

\newcommand{\var}{{\rm var}}

\begin{document}

\title{Number-unconstrained quantum sensing }

\newcommand{\myaffiliation}{\address}

\newcommand{\ICFO}
{
\myaffiliation{ICFO-Institut de Ciencies Fotoniques, The Barcelona Institute of Science and Technology, 08860 Castelldefels (Barcelona), Spain}}
\newcommand{\ICREA}
{
\myaffiliation{ICREA -- Instituci\'o Catalana de Recerca i Estudis
Avan\c{c}ats, 08015 Barcelona, Spain}
}
\newcommand{\ECNU}
{
\myaffiliation{East China Normal University, Shanghai 200062,
China}
}

\author{Morgan W. Mitchell}
\ICFO
\ICREA

\newcommand{\SINotMath}[2]{#1 #2}
\renewcommand{\micro}{[micro]}
\renewcommand{\SI}[2]{\ifmmode \mbox{\SINotMath{#1}{#2}} \else \SINotMath{#1}{#2} \fi}

\begin{abstract}
Quantum sensing is commonly described as a constrained optimization problem:  maximize the information gained about an unknown quantity using a limited number of particles. Important sensors including gravitational-wave interferometers and some atomic sensors do not appear to fit this description, because there is no  external constraint on particle number. Here we develop the theory of particle-number-unconstrained quantum sensing, and describe how optimal particle numbers emerge from the competition of particle-environment and particle-particle interactions. We apply the theory to optical probing of an atomic medium modeled as a resonant, saturable absorber, and observe the emergence of well-defined finite optima without external constraints. The results contradict some expectations from number-constrained quantum sensing, and show that probing with squeezed beams can give a large sensitivity advantage over classical strategies, when each is optimized for particle number. 
\end{abstract}

\maketitle



\newcommand{\unk}{x}
\newcommand{\unkGen}{{\cal X}}
\newcommand{\TNGen}{{\cal T}}
\newcommand{\noise}{n}
\newcommand{\el}{{\rm el.}}
\newcommand{\tech}{{\rm tech}}
\newcommand{\EN}{{\tilde s}}
\newcommand{\TN}{{\noise}}
\newcommand{\supzero}{^{(0)}}
\newcommand{\supstat}{^{({\rm stat})}}
\newcommand{\supsig}{^{({\rm stat})}}
\newcommand{\suptech}{^{({\rm tech})}}
\newcommand{\Ntot}{N_{\rm tot}}
\newcommand{\Nsat}{\nsat}
\newcommand{\FI}{{\cal I}}
\newcommand{\QFI}{\tilde{\cal I}}
\newcommand{\povm}{{\cal M}}
\newcommand{\povmEl}{M}
\newcommand{\povmOut}{m}
\newcommand{\adv}{{\cal A}}
\newcommand{\abs}{\xi}

\newcommand{\unkn}{{\cal X}}
\newcommand{\supin}{^{({\rm in})}}
\newcommand{\supout}{^{({\rm out})}}
\newcommand{\supn}{^{({n})}}
\newcommand{\supi}{^{({i})}}
\newcommand{\supj}{^{({j})}}
\newcommand{\supij}{^{({ij})}}
\newcommand{\supijk}{^{({ijk})}}


\section{Introduction}

Quantum sensing, the use of non-classical resources to improve the precision of interferometric measurements, is one of the earliest proposed applications of quantum optics \cite{CavesPRD1981} and a much-studied problem in quantum technology \cite{GiovannettiNPhot2011,DowlingJLT2015,PezzeARX2016,DegenARX2016,SimonJS2016,BraunARX2017}.  The core problem concerns the precision with which measurement on quantum systems can be used to estimate quantities that appear as classical parameters in the theory, for example time, displacement, rotation, and external fields. The practical ambition of quantum sensing is to enable precision instruments to reduce the effects of quantum statistical noise, e.g. shot noise. The theory of this field is remarkably wide-ranging, connecting statistics of parameter estimation \cite{Helstrom1976, Holevo1982} to the geometry of quantum states \cite{BraunsteinPRL1994} to entanglement in many-body systems \cite{SorensenPRL2001} to quantum information processing \cite{LeeJMO2002,GiovannettiPRL2006} to quantum non-locality \cite{TuraS2014,SchmiedS2016}.  

This essay aims to focus attention on the practical ambition of quantum sensing, and in particular on a large and technologically important class of problems that fall outside the scope of the more recent theoretical formulations of the problem. In particular, we study sensing problems for which $n$, the number of particles, is not a constraint or limiting resource, but is rather a free parameter chosen to optimize sensitivity. To readers familiar with quantum sensing, this may seem self-defeating.  The standard quantum limit  $\delta \unkn \propto 1/\sqrt{n}$, where $\delta \unkn$ is the uncertainty in a quantity to be sensed, trivially gives  $ \delta\unkn \rightarrow 0$ when $n$ is treated as a free parameter without limit. If the uncertainty can be reduced to zero through classical strategies, what purpose can exotic quantum strategies serve ? 

Upon closer inspection, we will see that this scenario, far from being trivial, is possibly the most important present-day scenario for practical quantum sensing.  It describes, for example, the GEO~600~\cite{LIGONP2011} and LIGO \cite{AasiNP2013Short} gravitational wave detectors when they were improved using squeezed light. These two real-world beneficiaries of quantum sensing, together with their analogues in spectroscopy \cite{PolzikPRL1992,LuciveroPRA2016,LuciveroARX2016} and atomic sensing \cite{WolfgrammPRL2010,HorromPRA2012,LuciveroRSI2014}, operate with $n$ chosen to optimize sensitivity, rather than externally constrained. As we will describe, the optimal $n$ emerges intrinsically, due to a competition between particle-environment and particle-particle interactions \cite{LIGONP2011,NapolitanoN2011,AasiNP2013Short,SewellPRX2014}, without need for an external constraint.  This  number-optimized scenario of interacting particles does not fit the assumptions of the problem of quantum sensing as usually defined \cite{GiovannettiNPhot2011,PezzeARX2016,DegenARX2016,SimonJS2016}, but nonetheless must be understood if we are to make progress with many real-world instruments. 

To make a first step in this direction, we describe the number-unconstrained sensing problem from the perspective of estimation theory, and employ the resulting formalism to study a widely-applicable model of spectroscopic sensing.  We observe that a finite optimal value for $n$ indeed emerges, due to saturation of the spectral features. The results contradict some expectations from number-constrained models of quantum sensing, while still indicating potentially large benefits from the technique. We find interesting scaling of Fisher information with the parameters of the material system, phase transitions between different optimal states, and a large advantage for number-optimized sensing with squeezed states over number-optimized sensing with classical states.

\section{Single-parameter sensing}


In a commonly-used formulation, the physical process within an interferometer is described by the transformation that converts an input state $\rho\supin$ to  an output state $\rho\supout$ 
\be
\label{eq:channel}
\rho\supout(\unkn) = L_\unkn(\rho\supin),
\ee
where $L_\unkn$ is a known quantum channel, i.e., a trace-preserving completely-positive map, parametrized by an unknown, continuous-valued classical parameter $\unkn$, the quantity we wish to know.  The quantum measurement \textit{per se} is described by a positive operator-valued measure (POVM)  $\povm \equiv \{\povmEl_i\}$ with elements $\povmEl_i$ and corresponding outcomes $\povmOut_i$, which by the Born rule 
have  probabilities
\be
\label{eq:BornRule}
P(\povmOut_i | \unkn) = {\rm Tr}[ \rho\supout(\unkn) \povmEl_i].
\ee
The information gained about $\unkn$ by this procedure can be quantified by various measures, most often the (classical) Fisher information, the expectation of the squared logarithmic derivative of the probability
\be
\label{eq:FIdef}
{\cal I}(\unkn) = E[ \{ \partial_\unkn  \ln P(\povmOut|\unkn)\}^2]
\ee 
which takes the form
\be
\label{eq:FIdef}
{\cal I}(\unkn) = \sum_i P(\povmOut_i | \unkn) \{\partial_\unkn  \ln P(\povmOut_i | \unkn)\}^2
\ee 
when $m$ is discrete, and 
\be
\label{eq:FIdef}
{\cal I}(\unkn) = \int d\povmOut\, P(\povmOut | \unkn) \{\partial_\unkn  \ln P(\povmOut | \unkn)\}^2
\ee 
when $m$ is continuous.  

From here on, we suppress the dependence of $\FI$ on $\unkn$, except where necessary to avoid confusion.  The Cramer-Rao bound places a lower limit on the mean square error of any estimator for $\unkn$ taking as input the measurement result $\povmOut$.  For a given $\povm$ and $\unkn$, the standard quantum limit (SQL) describes the maximal $\FI$ obtainable with separable  $\rho\supin$.   It is also common to optimize $\FI$ by choice of $\povm$ to find the quantum Fisher information  \cite{BraunsteinPRL1994}. 

%

\section{the standard problem in quantum metrology}
The above formulation is applicable to a wide variety of single-parameter estimation scenarios through the freedom in choosing $L_\unkn$.  It is very common, however, to specialize to channels of the form
\be
\label{eq:Lindep}
L_\unkn = l_\unkn^{(1)} \otimes l_\unkn^{(2)} \otimes \ldots \otimes l_\unkn^{(n)}
\ee
where $l_\unkn^{(i)}$ is a $\unkn$-parametrized quantum channel for the $i$th particle. This describes an independent interaction of each particle with the environment, without interaction among the particles participating in the sensing or other kinds of multi-particle effects. Such a transformation would be produced, for example, by a Hamiltonian
\be
H = \sum_i h\supi
\ee
where the $h\supi$ is a single-particle hamiltonian that acts on on the $i$th particle.  

A second  assumption is also nearly always adopted, which concerns not the nature of the physical system, but rather the resources available to the measurement, and hence the way of assigning value to different strategies for measuring $\unkn$.  In this assumption, $n$, the number of particles available to compose $\rho\supin$, or sometimes its average $\langle n \rangle = {\rm Tr}[n \rho\supin]$, is taken as a constraint. 
The optimization of $\FI$ subject to Eq. (\ref{eq:Lindep}), with $n$ or $\langle n \rangle$ as a constraint is the \textit{standard problem of quantum metrology} (SPQM).

Among the best-known SPQM results is the fact that  $\FI \propto n$ for non-entangled states, whereas there exist entangled states, i.e. Greenberger-Horne-Zeilinger states \cite{GreenbergerBook1989} or NooN states \cite{MitchellN2004} for which $\FI \propto n^2$ \cite{GiovannettiPRL2006}.  

\section{scaling and interactions}
\label{sec:Interactions}

Scaling of $\FI$ with $n$ has historically been given much importance.  
As just noted, non-entangled strategies scale as $\FI \propto n$, while entangled strategies can scale as $\FI' \propto n^2$, in the absence of interactions and technical noise. In this ideal scenario, entangled strategies will  beat non-entangled strategies for sufficiently large $n$, regardless of prefactors. This particular scaling argument has been shown to be of limited practical relevance, because it ignores the scaling of the cost of the entangled state, and the scaling of sensitivity to experimental imperfections \cite{EscherNP2011,Demkowicz-DobrzanskiNC2012,Demkowicz-dobrzanskiChapter2015}. Nevertheless, a similar scaling argument can help to understand number-unconstrained problems, and the connection to interactions among the sensing particles.

The particles used for sensing inevitably show some interaction, not just with the environment, but with each other. A Hamiltonian that describes a system of $n$ particles including $2$-body interactions, $3$-body interactions, and so forth is
\bea
H &=& H_1 + H_2 + H_3 + \ldots
\eea
where
\bea 
H_1 & \equiv & \sum_{i=1}^n h\supi_1, \,\,\,\,\,\,\,\,\,\,\,\,\,\,\,\,\,\,\,\,\,\,\,\,\,\,\,\,\,\,\,\,\,\,\,\,\,\,\,\,\,\, 
H_2  \equiv  \sum_{i=1}^n\sum_{j = 1}^{i-1} h_2\supij, \nonumber \\
H_3 & \equiv & \sum_{i=1}^n\sum_{j = 1}^{i-1} \sum_{k=1}^{j-1} h_3\supijk , \,\,\,\,\,\,\,\,\,\,\,\,\,\,\,\,\,\,\,\,\,\,\,\,\,\,  {\rm etc.}
\eea
Here $h_2\supij$ describes the two-body interaction of particle $i$ with $j$,  $h_3\supijk$ describes the three-body interaction of particles $i$, $j$ and $k$, and so forth.  The term $H_1$, which describes the interaction of single particles with their environment, gives rise to a quantum channel of the form given in Eq. (\ref{eq:Lindep}).  Neglecting the $H_{k>1}$ terms may appear natural in many situations, for example optical interferometry in vacuum where at first glance it would appear that any interaction among photons must be exceedingly weak.  As we shall see, this assumption is not as obvious as it might appear.

Any non-zero interactions, however small, force us to re-consider arguments such as the one presented in the first paragraph of this section.  Just as $\FI$ and $\FI'$ have different scalings with $n$, so do $H_1$ and $H_2$. However small $h_{2}$ may be, it is present $n(n-1)/2$ times in $H$, and for sufficiently large $n$ will surpass in importance the $h_1$ terms, which are present only $n$ times.  Which large-$n$ behaviour becomes important first?  Does $\FI'$ surpass $\FI$ before $H_2$ surpasses $H_1$ in its effect on $\rho$ ?  The answer  cannot be found through scaling arguments alone, rather we must consider details of the interacting system. 


%
%

%

%
%
%

\section{what sets $n$ in real sensing systems?}

We now consider, in a necessarily incomplete way, scenarios in which particle number is reasonably considered limited, or reasonably considered a free parameter. The particle number rarely reflects the number of particles actually available:  with a few exceptions such as francium \cite{GomezRPP2006} and anti-hydrogen \cite{AndresenN2010}, the particles used for sensing, typically photons, atoms, and electrons, are neither scarce nor expensive.  

\subsection{number-constrained}
\label{sec:Constrained}

Perhaps the earliest discussion of the number-limited condition is found in C. M. Caves' classic work ``Quantum-mechanical noise in an interferometer'' \cite{CavesPRD1981}. The abstract of that work offers a simple and historically-specific rationale:
\begin{quote}
The interferometers now being developed to detect gravitational waves work by measuring the relative positions of widely separated masses. Two fundamental sources of quantum-mechanical noise determine the sensitivity of such an interferometer: (i) fluctuations in number of output photons (photon-counting error) and (ii) fluctuations in radiation pressure on the masses (radiation-pressure error). Because of the low power of available continuous-wave lasers, the sensitivity of currently planned interferometers will be limited by photon-counting error. 
\end{quote}
In light of the low power, such an optical interferometer in vacuum could also be assumed to be a linear optical system, i.e., with no interaction among the photons.  Together, these considerations fit the SPQM picture, and given the importance of  Caves' article, one may suspect they directly inspired the SPQM. 

While the abstract correctly describes the situation in 1981, in that lasers of that time were of sufficiently low power as to make  photons a scarce resource, this situation did not persist, and the power of suitable lasers grew rapidly \cite{LIGONP2011}.  It is worth underlining that Caves' statement does not assume the number constraint as part of the definition of the sensing problem, but rather describes it as a contingency affecting the larger problem of improving instrumental sensitivity.

It is sometimes argued in single-photon quantum sensing that probing of delicate systems may place an externally-imposed limit on the number of probe photons \cite{BridaNPhot2010,DowlingJLT2015,JuffmannNComm2016,OnoNComm2013,WhittakerNJP2017}. Biological molecules, cells, and cell components have been  named as candidate delicate systems \cite{CrespiAPL2012, TaylorNPhot2013,TaylorThesis2015,TischlerSA2016} for this application.  An improved ratio of Fisher information to damage  has been demonstrated \cite{WolfgrammNPhot2013} in probing an atomic system. 

Last and certainly not least, trapped ions and atoms are usually constrained in particle number, at least for a particular trap with given dimensions.  Trapping mechanisms such as optical cooling function efficiently when the trap is empty or holds a small number of particles, but become less efficient as the trap ``fills up,'' producing an asymptotic approach to a maximum trap occupancy. Further steps such as evaporative cooling to produce colder samples are similarly limited by the initial number of trapped particles and the efficiency of the cooling.  Note that this number constraint is due to an interaction among the trapped particles.  So while one assumption of the SPQM, a limited particle number, is fulfilled, the other assumption, of non-interacting particles, does not immediately follow. For example, one of the challenges for ultra-cold atoms in implementing SPQM protocols is finding methods to ``turn off'' the naturally-present interactions \cite{FattoriPRL2008, MuesselPRL2014}.

\subsection{number-unconstrained}

There are other real-world scenarios in which particle number is not constrained in any meaningful way by considerations external to the sensor. Since the number of particles available is never infinite, this means that some other factor, internal to the sensor, must be limiting. A simple example, which we study in more detail in Section~\ref{sec:spectroscopy}, is laser spectroscopy, which typically is performed with laser power well below the maximum available.  Considering the signal-to-noise ratio (SNR), we can generically expect that the lowest powers will give poor SNR due to shot noise, and that due to saturation of the atomic or molecular transitions (a finite number of radiators can only radiate a finite field) the highest powers will also give poor SNR. Somewhere in between there will be an optimum.  If this optimum is below the available laser power, the scenario is not externally limited in particle number.  Note that it would not be adequate to describe this using the SPQM with the particle number set to its optimum, because the saturation, a nonlinear mechanism, arises from an interaction among the particles. 

With the important exception of cold, trapped atomic systems as noted in Section~\ref{sec:Constrained}, most atomic ensemble sensors are similarly unconstrained in particle number. In atomic vapors and gases the number density of atoms, and thus the number present in any finite-volume sensor, can be controlled by adjusting the vapor or gas pressure.  Molecules in solution, impurities in crystals, dopants in semiconductors and similar ensemble systems can also be adjusted in number density.  As with photons in spectroscopy, statistical fluctuations will be large at the lowest densities and interactions among the particles will bring in new physical effects at the highest densities, creating an optimum at an intermediate density. If this density is in fact reachable, the system is not subject to a relevant external number constraint. 

It is important to note that not all interactions reduce sensitivity.  Indeed, careful studies with atomic ensembles \cite{SewellPRX2014} have demonstrated scenarios in which Kerr optical nonlinearities, i.e. two-body interactions among photons \cite{NapolitanoNJP2010} are beneficial in quantum-noise limited probing of atoms \cite{NapolitanoN2011}.  Regarding atomic interactions, the most sensitive detectors for low-frequency magnetic fields \cite{KominisN2003,DangAPL2010} employ strong spin-exchange interactions to increase coherence times and thus sensitivity. The presence of such beneficial interactions imply a higher optimal $n$ and improved sensitivity, without fundamentally changing the nature of the unconstrained sensing problem. As shown in these same works, at still higher $n$ other, less-favourable interactions become important and create an optimum. 

Perhaps surprisingly, optical interferometry in vacuum can be in the number-unconstrained category.   When the GEO600 detector began using squeezed light \cite{LIGONP2011}, they described their motivation as follows:
\begin{quote}
The `classical' approach to improve the observatory's signal-to-shot-noise ratio is an increase of the circulating light power, as the signals produced by gravitational waves are proportional to the light power, whereas the shot noise is proportional to only the square root of the power. However, a higher light power leads to a thermal deformation of the sensitive interferometer optics and an increasing radiation pressure noise level, resulting in a practical upper limit for the optical light power applicable [ref]. Hence, further technologies must be considered to push the sensitivity beyond this limitation.
\end{quote}
Shortly thereafter, the LIGO H1 interferometer reported a similar application of squeezing \cite{AasiNP2013Short}, with a similar argument for its use:
\begin{quote}
To achieve the same improvement, a 64\% increase in the power stored in the arm cavities would have been necessary, but this power increase would be accompanied by the significant limitations of high-power operation [refs].
\end{quote}
The references,  \cite{PunturoCQG2010short} and \cite{HarryCQG2010,EvansPLA2010}, respectively, describe how parametric opto-mechanical effects can lead to increased noise in an optical cavity with a high circulating power.  The parametric interactions described are an optical nonlinearity, i.e., an interaction among photons, in which the radiation pressure from one photon can affect the behaviour of other photons, mediated by cavity mirror deformation. Evidently, it was not the scarcity of photons that, in 2011 and 2013, made quantum sensing a winning strategy for these instruments.  It was, rather, the non-obvious fact that quantum shot noise remained a limiting factor at the optimum defined by competition of linear and nonlinear effects. 

\section{number-optimized sensing and quantum limits}

It should be clear by now that important sensing instruments, including the few real-world examples of advantageous use of quantum resources, operate in a way not described by the SPQM.  In particular, these instruments have no important external constraint on particle number, and when the particle number is optimized, scaling leads naturally to a scenario in which an interaction among the particles counteracts the growth $\FI \propto n^d$ due to statistics, to give  an optimum described by $d\FI/dn = 0$.

\newcommand{\SQL}{{\rm SQL}}
\newcommand{\supSQL}{^{(\SQL)}}
\newcommand{\HL}{{\rm HL}}
\newcommand{\supHL}{^{(\HL)}}

For these optimized scenarios, the usual definitions of quantum limits, e.g. the SQL $\FI \propto n$, clearly do not apply. Ideas of quantum advantage tied to such SPQM definitions must also be revised.  For this purpose, the natural quantity to consider is $\FI\supSQL \equiv \max_{n, \rho\supin \in {\cal S}} \FI$, i.e. the largest Fisher information for any input state in the separable states ${\cal S}$, including free choice of $n$.  We may call this the number-optimized SQL.  
The quantum advantage can be defined as $\adv \equiv \FI'/\FI\supSQL$, where $\FI'$ is the Fisher information obtained with a given non-separable state.  Similarly,  $\FI$ optimized over all input states, including entangled ones, can be called the number-optimized Heisenberg limit $\FI\supHL$.  As we have seen, evaluation of these quantities will necessarily lead us to nonlinear models describing sensing with interacting particles \footnote{While nonlinear models have been discussed in the context of quantum sensing,  this work for the most part considers models that give scaling with $n$ that is monomial, i.e. $\FI \propto n^d$  \cite{BeltranPRA2005,BoixoPRL2007,BoixoPRL2008,NapolitanoNJP2010,NapolitanoN2011}, or exponential, i.e. $\FI \propto b^n$ \cite{RoyPRL2008}, in neither case exhibiting an optimum, and thus no different from the SPQM as regards the number-unconstrained scenario.  See \cite{SewellPRX2014} for a nonlinear sensing scenario exhibiting an optimum. }. 

\newcommand{\supres}{^{({\rm res})}}
\newcommand{\ares}{a\supres}

\section{number-optimized quantum sensing via spectroscopy}
\label{sec:spectroscopy}
We now present results on a simple case of nonlinear measurement, and show that it leads to a well-posed optimization problem in which the global solution, after optimizing the input state including the mean number of particles, retains quantum noise features that can be improved using quantum sensing techniques such as squeezing. 

We consider spectroscopy on a resonant medium, described in a refractive index model, so that the output field operator $a\supout$ is related to the input field operator $a\supin$ and an operator $\ares$ describing the absorption reservoir, assumed to be in its ground state.  We have 
\be
\label{eq:InOut}
a\supout = a\supin e^{i \phi}e^{- \abs} + \ares \left(1 - e^{-2\abs}\right)^{1/2},
\ee 
where the phase $\phi$ and attenuation $\abs$ are given by \cite{BoydBook2008}  
  \newcommand{\OD}{D_{\rm O}}
 \renewcommand{\OD}{{\rm OD}}
 \renewcommand{\OD}{T}
\newcommand{\nsat}{n_{\rm sat}}
 \bea
\phi + i \abs &=& \frac{\OD \gamma_0}{2} \frac{\Delta  + i \gamma }{\Delta^2 + \gamma^2}
\eea
where $\OD$ is the on-resonance optical depth, $\Delta$ is the detuning in angular frequency, and $\gamma_0$ is the unbroadened linewidth.  We include saturation of the medium by defining the power-broadened line-width 
\be 
\gamma^2 \equiv \gamma_0^2  \left(1+ \frac{\langle [a\supin]^\dagger a\supin \rangle}{{\nsat}}
\right),
\ee 
 $\nsat$ is the saturation photon number. We note that in this model $\gamma$ depends on the mean number of input photons $\langle [a\supin]^\dagger a\supin \rangle$, appropriate to conditions in which saturation takes place on a time-scale during which many copies of the state can be sent through the medium. This accurately describes many spectroscopy methods, and fits well with the use of Fisher information as a quantifier of performance: $\FI$ is related to estimator performance in the asymptotic regime, i.e., for  many uses of the state. 

It is convenient to normalize all frequencies by $\gamma_0$, defining $\bar{\Delta} \equiv \Delta/\gamma_0$, $\bar{\gamma} \equiv \gamma/\gamma_0$, to get 
\bea
\phi + i \abs &=& \frac{\OD}{2} \frac{\bar{\Delta}  + i  }{\bar{\Delta}^2 + \bar{\gamma}^2}
\eea


Using Eq. (\ref{eq:InOut}), we can now describe estimation of $\Delta$ or $\OD$ from measurement of the $X$ quadrature.  Estimation of $\Delta$ describes timekeeping in atomic clocks, in which a laser oscillator is referenced to an atomic line, and also to many atomic sensing techniques in which the frequency of a transition is measured to determine a quantity to be sensed, e.g. a magnetic field via the Zeeman shift of a transition. Estimation of $\OD$ describes quantification of the absorptive material, as in, e.g., imaging applications \cite{BridaNPhot2010}.  The combined loss and phase rotation gives the output signal 
\bea
\label{eq:MeanXout}
\langle X\supout \rangle &=&   \langle X_\phi\supin \rangle  e^{-\abs} 
\eea
where  $X_\phi \equiv 
a \exp[-i \phi] + a^\dagger \exp[i \phi]$ is a generalized quadrature, while the noise is described by
\bea
\label{eq:NoiseRIModel}
\var(X\supout ) &=& \var(X_\phi\supin  )e^{- 2\abs}  +  [1-e^{-2 \abs }], 
\eea
where the last term is the contribution from $\ares$.  These statistics fully determine the distribution of $X$ for gaussian input states. 

\newcommand{\rbs}{R_{\rm BS}}
\newcommand{\rtheta}{R_{\theta}}
\newcommand{\rphi}{R_{\phi}}
\newcommand{\R}{R}

A general gaussian input  is $D(\alpha)S(\zeta)|0\rangle$, where $|0\rangle$ indicates the vacuum, $D(\alpha) \equiv \exp[\alpha a^\dagger - \alpha^* a]$ displaces the state by $\alpha \equiv \R \exp[i \theta]$, and $S(\zeta) \equiv \exp[(\zeta^*a a-\zeta a^\dagger a^\dagger)/2]$ is the single-mode squeeze operator, with squeeze parameter $\zeta \equiv r \exp[2i  \psi]$.  The real parameters $\R,\theta,r$ and $\psi$ fully define the state.  The relevant statistics are
\bea
\langle X_\phi\supin \rangle &=& 2 \R \cos (\phi - \theta) \\
\var(X_\phi\supin ) & = & 
e^{-2r} \cos^2 (\phi-\psi) + e^{+2r} \sin^2 (\phi-\psi)\\
\langle a^\dagger a \rangle &=& 
\R^2 + \sinh^2 r.
\eea
We note that because the optical phase is explicitly included in the input state, it is sufficient to consider only the $X$ quadrature as a measurement.


\begin{figure}[t]
\begin{centering}
\hspace{-0.4cm}
{\includegraphics[width=0.9\columnwidth]{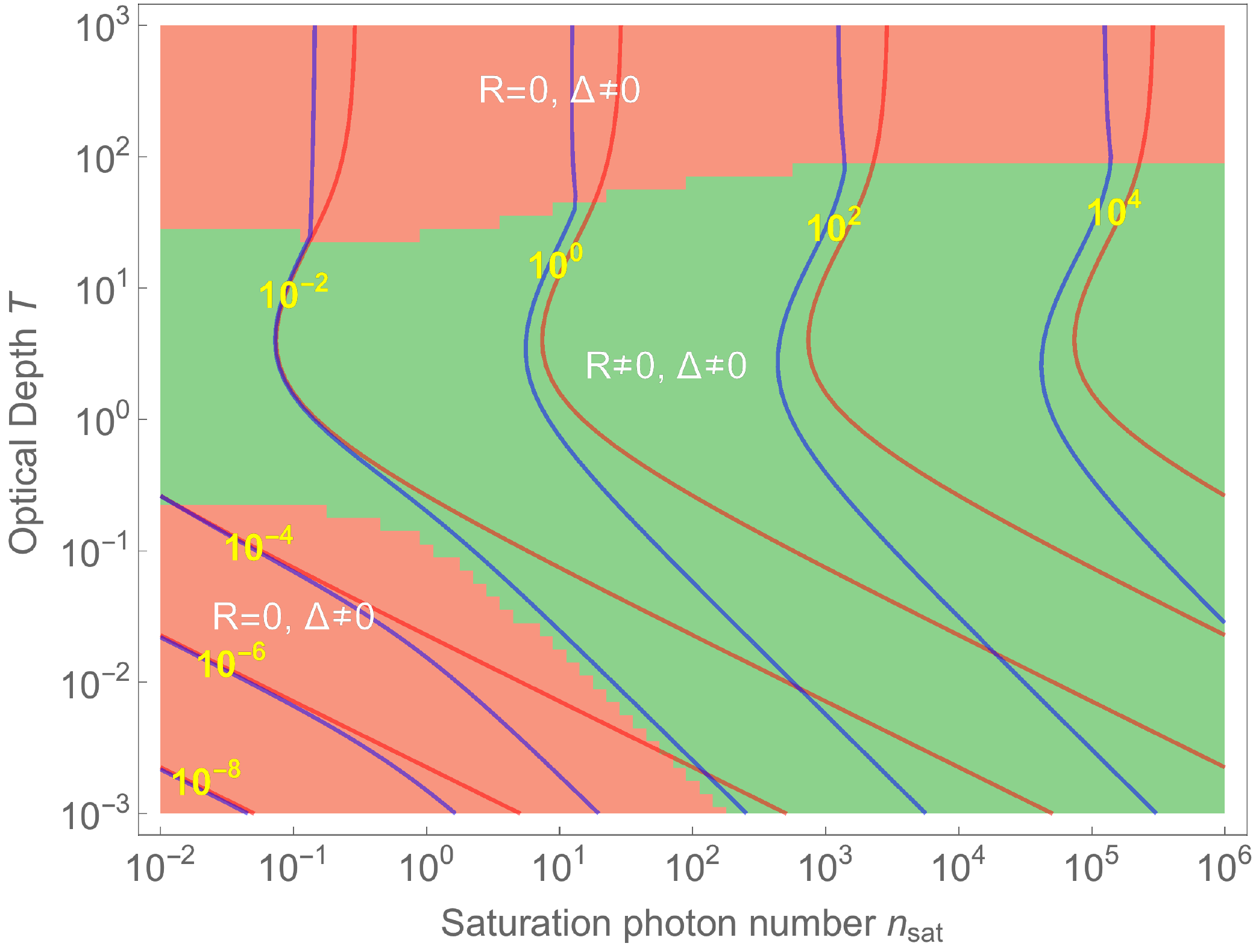}}
\end{centering}
\caption{Fully optimized Fisher information $\FI^{({\rm sq})}_\Delta$ (blue contours) and $\FI^{({\rm coh})}_\Delta$ (red contours) in sensing the detuning $\Delta$, in number-optimized probing of a saturable medium, as a function of the medium characteristics saturation photon number $n_{\rm sat}$ and on-resonance optical depth $\OD$.   $\FI^{({\rm sq})}_\Delta$ is the  Fisher information $\FI_\Delta$ obtainable by quadrature measurement on the output, maximized by choice of gaussian input state, including squeezed states. $\FI^{({\rm coh})}_\Delta$ is the same quantity optimized over only coherent input states, which defines the SQL.  Coloured regions indicate domains in which the optimal state is: off-resonance squeezed vacuum $R=0$, $\Delta \ne 0$ (red) 
or an off-resonance squeezed coherent state $R \ne 0$, $\Delta \ne 0$ (green).    }
\label{fig:AdvantageDelta}
\end{figure}

\begin{figure}[t]
\begin{centering}
\hspace{-0.4cm}
{\includegraphics[width=0.9\columnwidth]{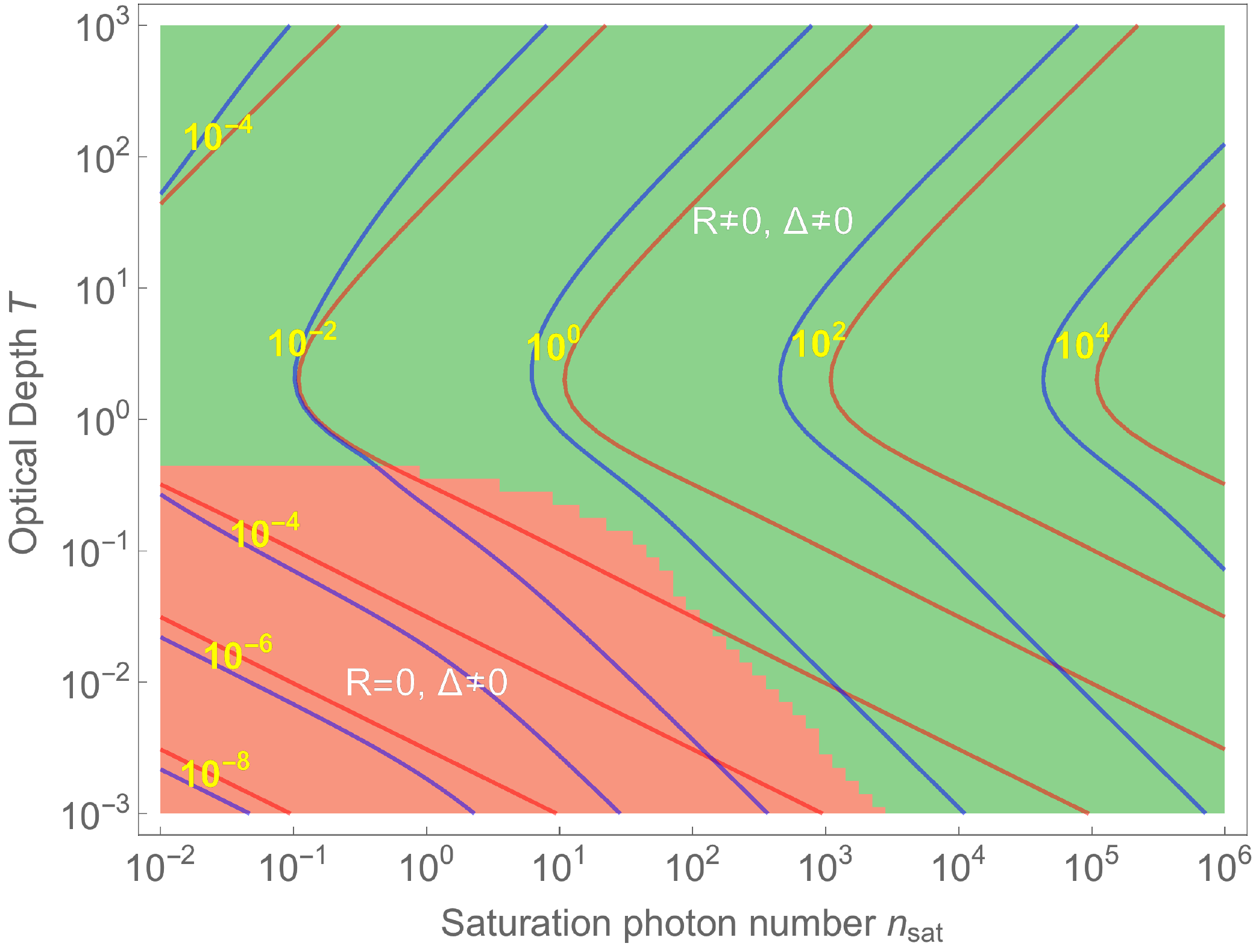}}
\end{centering}
\caption{Fully optimized Fisher information $\FI^{({\rm sq})}_\OD$ (blue contours) and $\FI^{({\rm coh})}_\OD$ (red contours) in sensing the optical depth $\OD$, in number-optimized probing of a saturable medium, as a function of the medium characteristics saturation photon number $n_{\rm sat}$ and on-resonance optical depth $\OD$.   $\FI^{({\rm sq})}_\OD$ is the  Fisher information $\FI_\OD$ obtainable by quadrature measurement on the output, maximized by choice of gaussian input state, including squeezed states. $\FI^{({\rm coh})}_\OD$ is the same quantity optimized over only coherent input states, which defines the SQL.  Coloured regions indicate domains in which the optimal state is: off-resonance squeezed vacuum $R=0$, $\Delta \ne 0$ (red) 
or an off-resonance squeezed coherent state $R \ne 0$, $\Delta \ne 0$ (green).    }
\label{fig:AdvantageDelta}
\end{figure}

\begin{figure}[t]
\begin{centering}
\hspace{-0.4cm}
{\includegraphics[width=0.9\columnwidth]{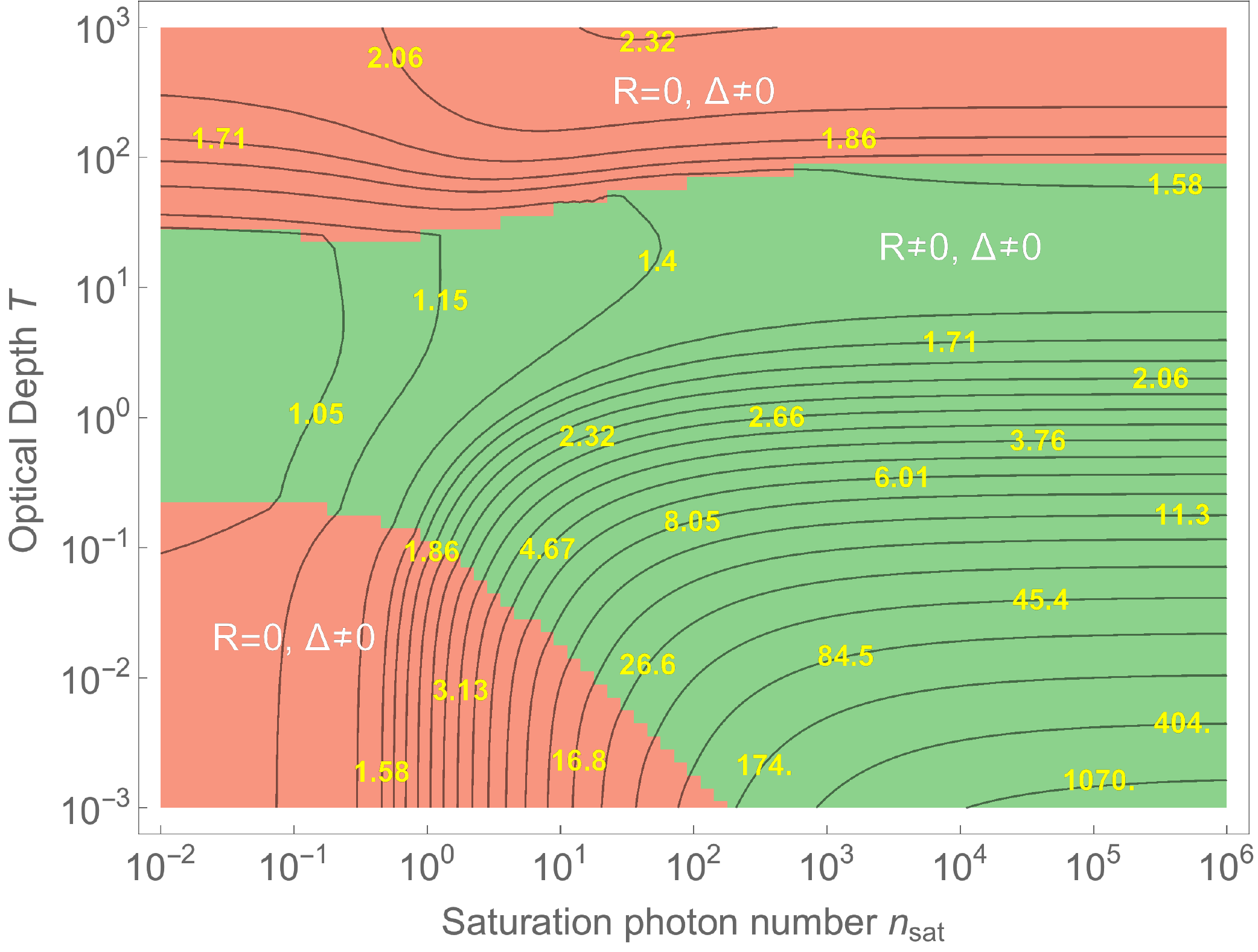}}
\end{centering}
\caption{Quantum advantage $\adv_\Delta \equiv \FI^{({\rm sq})}_\Delta/\FI^{({\rm coh})}_\Delta$ (labelled contours) in sensing the detuning $\Delta$, in number-optimized probing of a saturable medium, as a function of the medium characteristics saturation photon number $n_{\rm sat}$ and on-resonance optical depth $\OD$.   $\FI^{({\rm sq})}_\Delta$ is the  Fisher information $\FI_\Delta$ obtainable by quadrature measurement on the output, maximized by choice of gaussian input state, including squeezed states. $\FI^{({\rm coh})}_\Delta$ is the same quantity optimized over only coherent input states, which defines the SQL.  Coloured regions indicate domains in which the optimal state is: resonant squeezed vacuum $R=0$, $\Delta \ne 0$ (red) 
or an off-resonance squeezed coherent state  $R \ne 0$, $\Delta \ne 0$ (green).    }
\label{fig:AdvantageDelta}
\end{figure}

\begin{figure}[t]
\begin{centering}
\hspace{-0.4cm}
{\includegraphics[width=0.9\columnwidth]{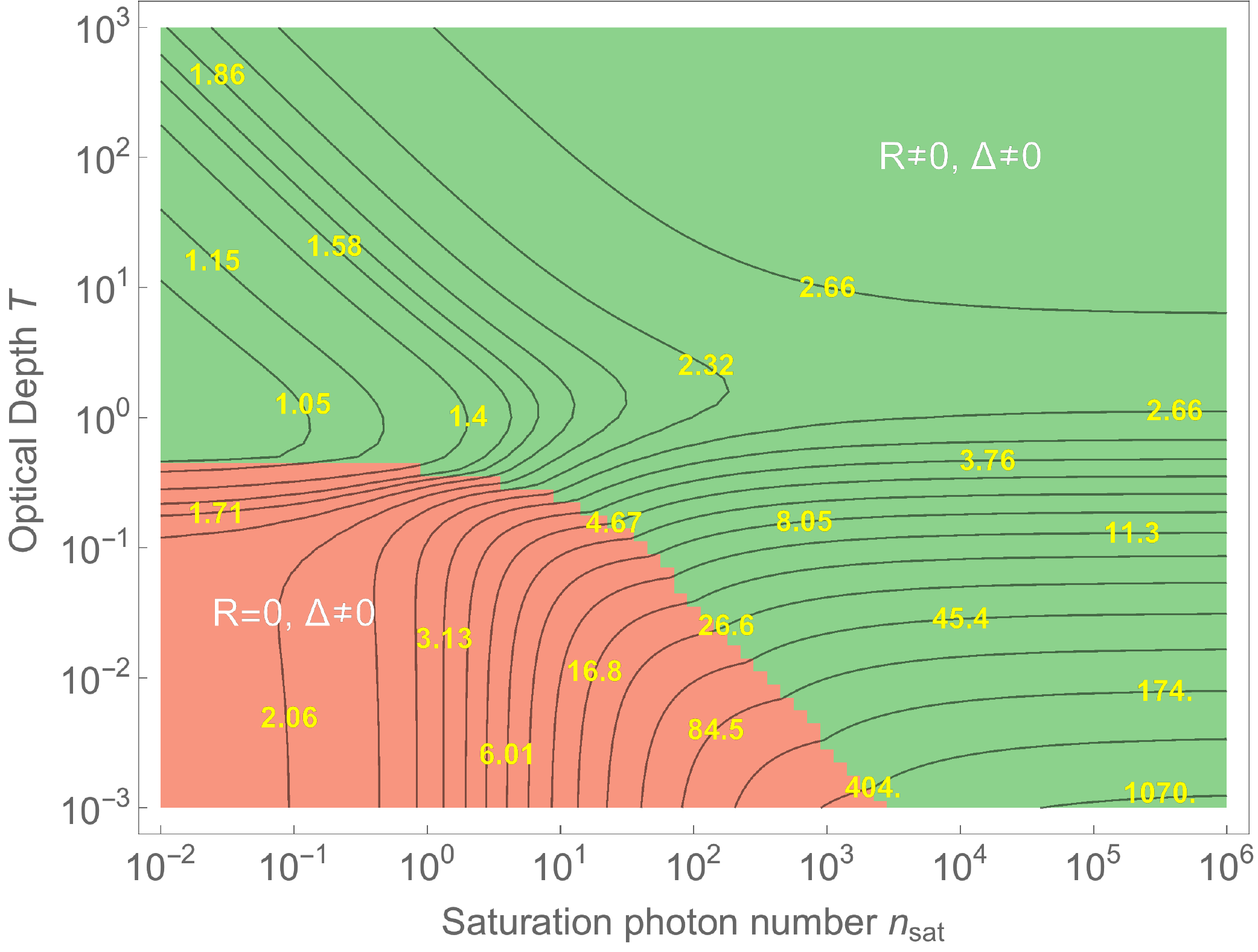}}
\end{centering}
\caption{Quantum advantage $\adv_\OD \equiv \FI^{({\rm sq})}_\OD/\FI^{({\rm coh})}_\OD$ (labelled contours), in sensing the  on-resonance optical depth $\OD$.  Representation is the same as in Fig. \ref{fig:AdvantageDelta}.  }
\label{fig:AdvantageOD}
\end{figure}


\newcommand{\FIParam}{{\cal Y}}
\renewcommand{\FIParam}{\unkn}
For a gaussian distribution with mean $\mu$ and variance $v$, both  depending on a parameter $\FIParam$, the Fisher information is  
\be
\FI_\FIParam = \frac{1}{v}\left(\frac{\partial \mu}{\partial \FIParam} \right)^2 +\frac{1}{2v^2}\left(\frac{\partial v}{\partial \FIParam} \right)^2.
\ee
We apply this to measurement of $X$ using $\mu \equiv \langle X \rangle$ and  $v \equiv \var( X)$ from Eqs. (\ref{eq:MeanXout}) and (\ref{eq:NoiseRIModel}), respectively, for parameters of interest $\FIParam \in \{\Delta,\OD\}$, as appropriate for the measurements of resonance frequency and material density, respectively. 


For given $T$ and $\nsat$, which parametrize the characteristics of the system rather than the probing, we can identify the optimal gaussian-state measurement strategy by maximizing $\FI$ with respect to $\R,\theta,r,\psi$ and $\Delta$, defining the optimum as $\FI^{(\rm sq)}$.  We underline that this describes the best possible measurement using gaussian states, including the freedom to choose the mean number of particles.   By setting $r = 0$ and optimizing  $\FI$ by choice of $\R,\theta$ and $\Delta$,  we find $\FI^{(\rm coh)}$, which describes the SQL, i.e. the best possible measurement using classical input states. The number-optimized quantum advantage is $\adv_\FIParam \equiv \FI^{({\rm sq})}_\FIParam/\FI^{({\rm coh})}_\FIParam$. 

Numerical results are shown in Figs. \ref{fig:AdvantageDelta} and  \ref{fig:AdvantageOD}.  For estimates of $\Delta$, we find that squeezing always benefits the Fisher information, for some parameter regimes by a large amount, with an increasing benefit for small $\OD$ and large $\nsat$.  For $\nsat >1$, the advantage scales as $\adv \sim \nsat$ until saturating at a value $\adv \sim 1/T$.  Perhaps counter-intuitively, the quantum advantage persists even at large $T$, approaching a constant value $\adv_\Delta \approx 2$ for $\nsat\lesssim10$ and somewhat higher values for larger $\nsat$.  Many of these same observations hold for estimation of $\OD$, with the notable exception that $\adv_{\OD} \approx 1$ in the region $\OD\sim 1$, $\nsat < 1/10$, indicating that squeezing provides little advantage here.  Curiously, this region is in the easily-saturated regime often proposed as promising for application of non-classical states \cite{BridaNPhot2010, DowlingJLT2015, JuffmannNComm2016, OnoNComm2013, WhittakerNJP2017}. In contrast, $\adv_\OD$ appears to grow without limit as $\OD$ moves away from $1$, either to larger or smaller values.  Finally, it is interesting to note that the character of the optimal state makes abrupt transitions, with squeezed coherent states taking over from squeezed vacuum states with increasing $\nsat$, and in the case of $\Delta$ estimation, $\Delta=0$ states winning for large $T$ and $\nsat$. 

\section{conclusions and outlook}

It is clear from these results on probing of saturable resonant media that a number-unconstrained approach to quantum sensing can lead to quite different conclusions than does the standard, number-limited description of quantum sensing (the SPQM). Most evidently, it is the interaction among sensing particles, something completely absent from the SPQM, that plays the strongest role in determining the advantage achievable with quantum resources.  Other notable differences include the prediction that nonclassical states do not always benefit imaging of sensitive materials easily damaged by an optical probe, and the observation of phase transitions between optimal sensing states.  It should be stressed that, by the nature of the unconstrained sensing problem, the results are necessarily specific to the material model, and in particular to its nonlinear behaviour. The model we have chosen describes a wide variety of optical measurements, both in spectroscopy and in optically-detected atomic sensing, e.g. in optical magnetometry. Further work is needed to understand other models. In the future, it will be interesting to extend this treatment to include also atomic quantum noise, given that in many real-world sensors the atom number is similarly unconstrained. It will also be interesting to study the more elaborate protocols often employed in real instruments, e.g. Ramsey sequences or measure-evolve-measure protocols \cite{SewellNP2013}.  We expect the study of such number-unconstrained sensing problems will yield new insights into quantum sensing for a large class of practical instruments, which includes gravitational-wave detectors, spectroscopic sensors, imaging systems, and atomic sensors.


%
%
%
%
%
%
%

\section*{Acknowledgements}

We thank I. Bouchoule, R. Jimenez-Martinez,  R. Sewell and C. Westbrook for helpful comments.  This work was supported by European Research Council (ERC) projects AQUMET (280169) and ERIDIAN (713682); European Union QUIC (641122); Ministerio de Econom'a y Competitividad (MINECO) Severo Ochoa programme (SEV-2015-0522) and projects MAQRO (Ref. FIS2015-68039-P), XPLICA  (FIS2014-62181-EXP); Ag\`{e}ncia de Gesti\'{o} d'Ajuts Universitaris i de Recerca (AGAUR) project (2014-SGR-1295) and Fundaci\'{o} Privada CELLEX.

\section*{References}
\bibliography{./BigBib170101}

\end{document}